\documentclass[useAMS,usenatbib]{mn2e}
\usepackage{times}
\usepackage{rotate}
\usepackage{graphicx}
\usepackage{graphics}
\input{epsf}
\newif\ifAMStwofonts
\AMStwofontstrue
\usepackage{relsize,txfonts}
\usepackage{color}

\title
{Excess AGN Activity in the $z=2.30$ Protocluster in HS 1700+64}

\author[J. A. Digby-North et al.]{J.~A.~Digby-North$^{1,}$\thanks{j.digby-north07@imperial.ac.uk}, K.~Nandra$^{1}$, E.~S.~Laird$^{1}$, C.~C.~Steidel$^{2}$, A.~Georgakakis$^{3}$, \newauthor M.~Bogosavljevi\'{c}$^{2}$, 
 D.~K.~Erb$^{4}$, A.~E.~Shapley$^{5}$, N.~A.~Reddy$^{6}$ and J.~Aird$^{7}$\\
$^1$Astrophysics Group, Imperial College London, Blackett Laboratory,
Prince Consort Road, London SW7 2AW, UK\\
$^2$California Institute of Technology, MS 105-24, Pasadena, CA 91125, USA\\
$^3$National Observatory of Athens, V. Paulou and I. Metaxa, 11532, Greece\\
$^4$University of California Santa Barbara, Department of Physics, Santa Barbara, CA 93106, USA\\
$^5$Department of Physics and Astronomy, 430 Portola Plaza, University of California at Los Angeles, Los Angeles, CA 90025, USA\\ 
$^6$National Optical Astronomy Observatory, 950 N Cherry Ave, Tucson, AZ 85719, USA \\
$^7$Department of Physics, University of California, San Diego, CA 92093, USA\\}

\begin{document}

\date{06 May 2010}

\pagerange{\pageref{firstpage}--\pageref{lastpage}} \pubyear{2010}

\maketitle
\label{firstpage}

\begin{abstract}
We present the results of spectroscopic, narrow--band and X-ray observations of a $z=$ 2.30 protocluster in the field of the QSO HS 1700+643.  Using a sample of BX/MD galaxies, which are selected to be at $z\sim$ 2.2--2.7 by their rest-frame ultraviolet colours, we find that there are 5 protocluster AGN which have been identified by characteristic emission-lines in their optical/near-IR spectra; this represents an enhancement over the field significant at $\sim$98.5~per cent confidence.  Using a $\sim$200~ks {\it Chandra}/ACIS-I observation of this field we detect a total of 161 X-ray point sources to a Poissonian false-probability limit of $4\times10^{-6}$ and identify 8 of these with BX/MD galaxies.  Two of these are spectroscopically confirmed protocluster members and are also classified as emission-line AGN. When compared to a similarly selected field sample the analysis indicates this is also evidence for an enhancement of X-ray selected BX/MD AGN over the field, significant at $\sim$99~per cent confidence.  Deep Ly$\alpha$ narrow-band imaging reveals that a total of 4/123 Ly$\alpha$ emitters (LAEs) are found to be associated with X-ray sources, with two of these confirmed protocluster members and one highly likely member. We do not find a significant enhancement of AGN activity in this LAE sample over that of the field (result significant at only 87~per cent confidence).  The X-ray emitting AGN fractions for the BX/MD and LAE samples are found to be $6.9_{-4.4}^{+9.2}$ and $2.9_{-1.6}^{+2.9}$~per cent, respectively, for protocluster AGN with $L_{\rm 2-10 keV}\geq4.6\times10^{43}$~erg~s$^{-1}$ at $z=$ 2.30.  These findings are similar to results from the $z=$ 3.09 protocluster in the SSA 22 field found by Lehmer et al. (2009), in that both suggest AGN activity is favoured in dense environments at $z>2$.
\end{abstract}


\section{Introduction}
The effect of local environment on galaxy formation and evolution has been the subject of many recent studies. Observations of local clusters show a strong dependence of stellar age on the local environment ({\rm e.g.} Smith et al. 2009); the oldest and most evolved galaxies are found in the central regions whereas the younger actively star-forming galaxies are common on the outskirts, in lower density regions.  Studies of higher redshift ($z\sim 1$) clusters have also shown that the stars in massive early-type galaxies must have formed at $z>2$ ({\rm e.g.} van Dokkum $\&$ Stanford 2003) with the galaxies mainly passively evolving since that time; this is consistent with results from Thomas et al. (2005) who find that massive galaxies in high density environments are expected to have formed the bulk of their stars at much earlier times than their counterparts in the `field'. These observations are in general agreement with models which predict the accelerated formation of galaxies in overdense environments ({\rm e.g.} Kauffmann 1996).  

There is also evidence that Active Galactic Nuclei (AGN) may be more prevalent in high density regions. Studies of X-ray selected AGN show that they are strongly clustered at $z\sim1$ ({\rm e.g.} Miyaji et al. 2007 and references therein), like the evolved, red galaxies which host them (Coil et al. 2009). Many also appear to reside in group environments (Georgakakis et al. 2008a). In the vicinity of clusters, at least at $z<1$, the X-ray source density appears enhanced ({\rm e.g.} Cappi et al. 2001; Martini et al. 2002;  Cappelluti et al. 2005; Gilmour, Best \& Almaini 2009).  Spectroscopic and X-ray observations of low redshift clusters by Martini et al. (2006) also show that a substantial number of their cluster galaxies exhibit AGN activity and indicate that differing AGN identification methods result in significant differences between the measured AGN fraction in these environments, and therefore its subsequent relation to that of the field.   As the galaxy number density is much higher in clusters it is reasonable to also expect a higher interaction and merger rate in these regions, and one hypothesis for the triggering of AGN activity at high-$z$ also involves mergers ({\rm e.g.} Springel, Di Matteo \& Hernquist 2005).  In this scenario, we might therefore expect to observe an enhancement in the fraction of galaxies exhibiting signs of AGN activity in high-$z$ overdense regions when compared to the field -- especially as these pre-virialized structures may be the ideal place for efficient mergers of galaxies to result in increased AGN activity ({\rm e.g.} van Breukelen et al. 2009).  Also, recent studies of the evolution of the X-ray AGN fraction in clusters of galaxies (for $z<1.3$) show a significant decrease in the observed fraction of cluster galaxies hosting AGN over cosmic time (Eastman et al. 2007; Martini, Sivakoff \& Mulchaey 2009), and so the limited observations of high redshift ($z>2$) overdense regions become increasingly important in determining the situation in the early universe.

\begin{figure}
{\scalebox{0.31}
{\includegraphics{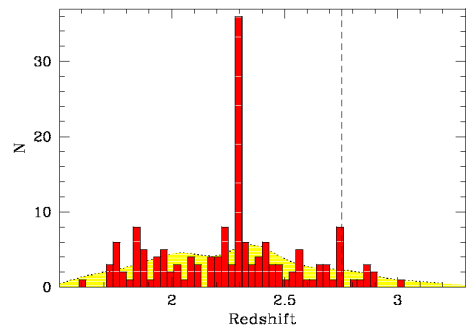}}}
\centering
\caption{Redshift distribution for the 178 spectroscopically confirmed BX/MD galaxies with $z>1.5$ in the HS 1700+64 field.  The background shaded curve (yellow) is the expected distribution of redshifts when drawn from a randomly selected sample of BX/MD galaxies.  The vertical dashed line indicates the redshift of the QSO HS 1700+643.  An overdensity of galaxies at $z=$ 2.3 is observed (Steidel et al. 2005).}
\label{fig:zspike}
\end{figure}

Studying the precursors of local massive clusters (``protoclusters") at $z>2$ might thus reveal valuable information regarding the earliest (and most active) stages of cluster formation, the progenitors of local massive cluster galaxies and how nuclear activity is related to the local environment.  The highest redshift protoclusters have been found by searching for an excess of Ly$\alpha$ or H$\alpha$ emitting galaxies near massive high-$z$ radio galaxies ({\rm e.g.} Venemans et al. 2007), or during the course of spectroscopic surveys for high redshift galaxies ({\rm e.g.} Steidel et al. 1998). There are relatively few robustly established overdensities; the $z=$ 3.09 protocluster in the SSA 22 field is one such structure (Steidel et al. 1998).  This field was found to have both an overdensity of Lyman Break Galaxies (LBGs) and LAEs.  Lehmer et al. (2009) studied the X-ray emission from the LBGs and LAEs in the protocluster, and when compared to the field (at $z\sim 3$) they claim suggestive evidence for an enhancement (for log $L_{\rm 8-32 keV}\gtrsim 43.5$~erg~s$^{-1}$) by a factor $\sim$5 and $\sim$7 for LBGs and LAEs, respectively, significant at a combined level of $\sim$95~per cent.  The HS 1700+64 field contains a similar overdensity of galaxies at $z=2.300 \pm 0.015$ (Steidel et al. 2005, hereafter S05; Fig.~\ref{fig:zspike}).  This structure is predicted to virialize by $z\sim$ 0 with a cluster-like mass scale of $\sim$1.4$\times$10$^{15}{\rm M}_{\tiny{\sun}}$.  Due to the available data in this field, the HS 1700 protocluster is an excellent laboratory for discerning AGN (and host galaxy) properties in a very high density large-scale structure environment. Here we use a combination of optical/near-IR spectroscopy, sensitive {\it Chandra} observations and narrow-band imaging of the protocluster with the aim of testing these ideas.

Thoughout this work a standard, flat $\Lambda$CDM cosmology with $\Omega_{\Lambda}=$ 0.73 and H$_0=$ 70~km~s$^{-1}$~Mpc$^{-1}$ is assumed.

\section{Observations}

\subsection{Optical Imaging and Spectroscopy}
The Q1700 field ($15.3 \times 15.3$~arcmin) was observed in the {\it U$_n$}, {\it G} and $\cal R$ bands with the Prime Focus Imager on the William Hershel 4.2 m telescope (Steidel et al. 2004).  In the Q1700 field, star-forming galaxies in the foreground to the $z=$ 2.72 QSO HS 1700+643 were targetted using the `BX' optical photometric criteria (Adelberger et al. 2004; Steidel et al. 2004; $<z>$ $=$ 2.20 $\pm$ 0.32) and the `MD' criteria of Steidel et al. (2003; $<z>$ $=$ 2.73 $\pm$ 0.27).  Follow-up spectroscopy (probing rest-frame UV wavelengths) was taken with the LRIS-B instrument on the Keck I telescope.  There are also 19 H$\alpha$ redshifts available in the Q1700 field (Erb et al. 2006) from the near-infrared spectrograph NIRSPEC on the Keck II telescope; H$\alpha$ falls in the $K_{s}$--band at $z=2.3$ and so this was used in the selection of H$\alpha$ emitters in the Q1700 field.  

In total, there are 1710 photometric BX and MD candidate galaxies to $\cal R=$ 25.5; 1472 are identified as BX and 238 as MD.  We have secure spectroscopic redshifts for 213 ($\sim$12~per cent) of these; this sample consists of 175 BX galaxies ($\sim$12~per cent of the photometric BX sample) and 38 MD galaxies ($\sim$16~per cent of the photometric MD sample).  Of the spectroscopic BX and MD samples, 147 BX galaxies ($\sim$84~per cent) and 31 MD galaxies ($\sim$82~per cent) have $z>$ 1.5; the mean redshift of this BX/MD sample is $<z>$ $=$ 2.28 $\pm$ 0.31.  Regarding protocluster membership, there are 46 galaxies (40 BX and 6 MD) whose spectroscopic redshifts place them within the bounds of the structure at $z=$ 2.30 $\pm$ 0.04; this is $\sim$22~per cent of the spectroscopic BX/MD selected sample.  Objects contaminating the spectroscopic BX/MD sample were low redshift ($z<$ 1.5) star-forming galaxies ($\sim$12~per cent) and stars ($\sim$4~per cent).  The spectroscopic sample discussed here is larger than that of S05 due to subsequent identifications with Keck/LRIS-B. 

\subsection{Chandra Observations and Source Detection}
The HS 1700+64 field was observed by {\it Chandra} in November 2007, using the ACIS-I instrument ($16.9 \times 16.9$~arcmin).  There were eight individual pointings totalling $\sim$200~ks of exposure; the observation numbers (ObsIDs) and other basic information are detailed in Table~\ref{tab:obs}.  The data reduction was completed with the {\it Chandra} X-ray Center (CXC) {\it Chandra} Interactive Analysis of Observations (CIAO) data analysis software, version 3.4, and the {\it Chandra} calibration database (CALDB), version 3.3.

Each ObsID was reduced separately in a similar manner to that described by Laird et al. (2009), hereafter L09. Briefly, the raw level 1 event files were first corrected for known systematic astrometric offsets; hot/bad pixels and cosmic ray afterglows were also removed.  Level 2 (filtered) event files were created by applying the charge-transfer inefficiency (CTI) calibration and time-dependent gain correction, then removing the ACIS pixel randomization and applying the ACIS particle background cleaning algorithm.  Standard screening criteria were applied to the observations (using the standard `ASCA' grade set), with Good Time Intervals (GTI) applied; this resulted in the removal of $\sim$2.4~ks of exposure time.  The observations were then aligned using the CIAO task {\it align\_evt} and merged to create level 2 event files and merged images; these were created in the full (0.5--7 keV), soft (0.5--2 keV), hard (2--7 keV) and ultrahard (4--7 keV) bands.  Weighted exposure maps were created for each ObsID, with merged exposure maps being created in a similar manner to the images.  

Source detection and photometry also follows L09.  Briefly, candidate sources are initially identified in the above energy bands via use of the CIAO algorithm {\smaller{WAVDETECT}}, running at a significance threshold of $10^{-4}$. Point Spread Functions (PSFs) were calculated (as described in L09) for each of the sources; the 70~per cent Encircled Energy Fraction (EEF) PSF was used to extract the X-ray counts and determine their significance compared to the local background.  A Poissonian false-probability threshold of $4\times10^{-6}$ was applied (see Nandra et al. 2005 for details) which resulted in 161 band-merged source detections -- each of these has a Poissonian false-probability value which is below the threshold in at least one of the four energy bands. At the aimpoint of the ACIS-I $\sim$200~ks observation, the hard-band (2--10 keV) X-ray luminosity limit is approximately $L_{\rm 2-10 keV}\sim$ 6.5$\times10^{42}$~erg~s$^{-1}$ at $z=$ 2.30.  Any X-ray point sources we detect above this limit are assumed to be AGN based on the energetics involved ({\rm e.g.} Steffen et al. 2003).

\subsection{Narrow-band Imaging}
Deep (22.5hr integration) Ly$\alpha$ narrow-band imaging was obtained using a custom built narrow-band filter centred at 4010$\rm \AA$ with FWHM $\sim 90\rm \AA$ with the Large Format Camera on the Palomar 200in Hale telescope (P200).  The images reach narrow-band depths of NB $<$ 25.5 (AB) and cover the entire WHT field.  As the 4010$\rm \AA$ filter bandpass falls between the {\it U$_n$} and {\it G} broad-band filters, an average of the {\it U$_n$} and {\it G} measurements was used to predict the continuum level at 4010$\rm \AA$.  Ly$\alpha$ excess candidates were then selected based on the difference between the NB4010 flux and that predicted for the continuum from the broad-band filters.  The criterion required for selection was $\rm NB4010-UG\le-0.75$, where $\rm NB4010$ and $\rm UG$ refer to the magnitudes of the narrow-band and continuum measurements, respectively. This corresponds to a requirement of $\rm EW_{obs}>90\rm \AA$.   There are 123 narrow-band excess objects satisfying these criteria in the HS 1700+64 field, 33 of which are spectroscopically confirmed.  The fraction of spectroscopic LAEs which lie at the protocluster redshift is $\sim$88~per cent.  Of this LAE sample, 19 objects also have a BX/MD counterpart within a 1.5~arcsec radius of the LAE position.  We also note that the sample of LAEs includes 6 Ly$\alpha$ `blobs' (LABs; {\rm e.g.} Smith \& Jarvis 2007).  These LABs were selected in the same way as the LAEs, but with the additional requirement that the isophotal Ly$\alpha$ area was larger than 50 sq arcsec.

We also have deep (21hr) H$\alpha$ narrow-band imaging in this field, obtained with the P200.  The imaging covers a central $9 \times 9$~arcmin region of the field, roughly centred on the background QSO HS 1700+643.  The Bracket gamma filter centred at 2.17$\micron$ with FWHM $=297\rm \AA$ was used to select the candidate H$\alpha$ emitters.  The continuum estimate was obtained from the $K_{s}$--band magnitude.  Selection then required $\rm Br\gamma - K_{s}\le-0.6$, where $\rm Br\gamma$ and $\rm K_{s}$ are in magnitudes.  This translates into $\rm EW_{obs}>220\rm \AA$.  There are 62 H$\alpha$ emitters (HAEs) in the field which fulfil our selection criteria.  Of these 23 have confirmed redshifts, with 12 that lie within the bounds of the structure at $z=2.30$; 18 HAEs have a BX/MD counterpart within 1.5~arcsec.

\begin{figure*}
{\scalebox{0.5}
{\includegraphics{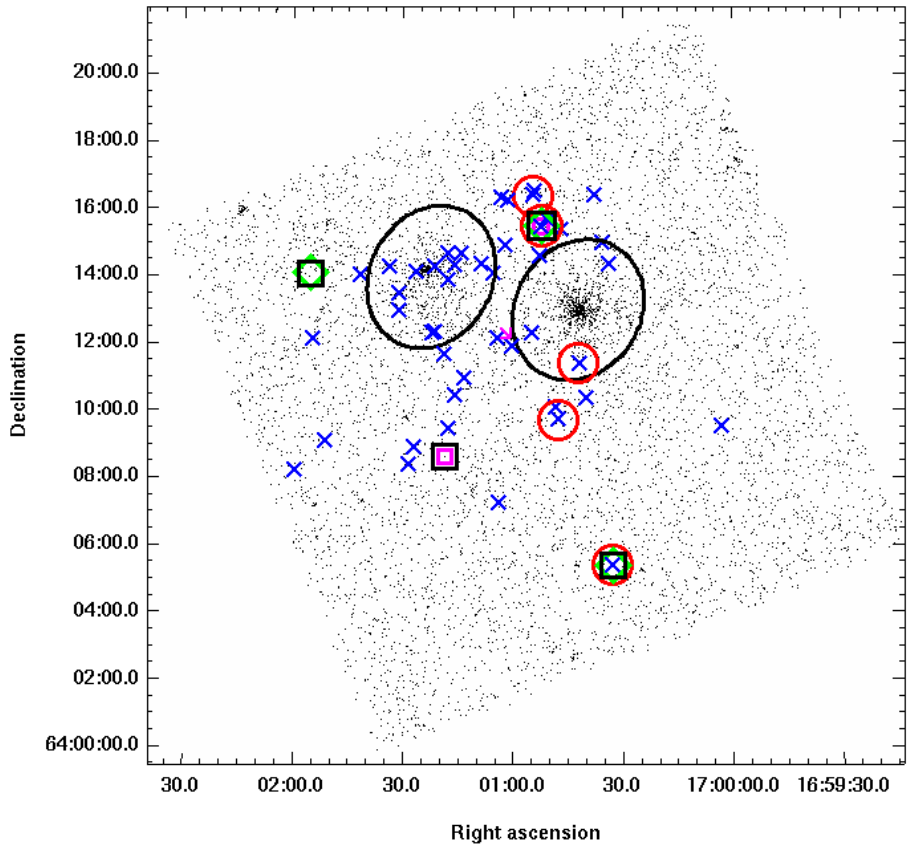}}}
\centering
\caption{{\it Chandra} full band image of the HS 1700+64 field.  Black boxes represent X-ray detections associated with the protocluster; spectroscopically confirmed BX/MD protocluster members are denoted with a blue cross; red circles show protocluster emission-line AGN; X-ray detected protocluster Ly$\alpha$ objects are the green diamonds; small magenta squares denote protocluster H$\alpha$ narrow-band detections associated with X-ray sources; the ellipses centred on the two foreground clusters are the regions excluded from the X-ray analysis and the magenta arrow shows the position of the background QSO HS 1700+643.}
\label{fig:LAEdist}
\end{figure*}

\begin{table}
\centering
\caption{Log of Chandra observations. 
(1): Observation ID number; (2): Nominal right ascension of pointing; (3): Nominal declination of pointing; (4): Start date and time of observation; (5): Filtered exposure time}
{\smaller
\begin{tabular}{cccccc}
\hline
ObsID & RA  & Dec  & Start Time/Date & Filt. Exp\\
 & (J2000) & (J2000) & (UT) & (ks)\\
(1) & (2) & (3) & (4) & (5)\\
\hline 
8032 & 17 00 55.00  & +64 11 26.70 & 2007-11-12 02:44:29 & 31.0\\
8033 & 17 00 55.00  & +64 11 26.70 & 2007-11-20 01:12:58 & 29.7\\
9756 & 17 00 55.00  & +64 11 26.70 & 2007-11-14 17:59:55 & 32.3\\
9757 & 17 00 55.00  & +64 11 26.70 & 2007-11-13 17:47:55 & 20.8\\
9758 & 17 00 55.00  & +64 11 26.60 & 2007-11-16 09:09:31 & 23.4\\
9759 & 17 00 55.00  & +64 11 26.80 & 2007-11-17 09:08:16 & 31.2\\
9760 & 17 00 55.10  & +64 11 26.70 & 2007-11-19 01:53:05 & 17.0\\
9767 & 17 00 55.00  & +64 11 26.90 & 2007-11-21 09:32:28 & 9.0\\
\hline
\label{tab:obs}
\end{tabular}
}
\end{table}

\section{Results}

\subsection{Emission-line AGN}
There are a number of BX/MD selected AGN in HS 1700+64 that have been identified from their rest-frame UV and/or rest-frame optical spectra.  Of the 8 spectroscopically identified AGN in the field with $z_{\rm spec}>1.4$, 5 are within $\delta z=\pm0.04$ of the protocluster central redshift (see Table~\ref{tab:properties}). These objects have been identified as AGN based on the presence of high ionization emission lines (${\rm e.g.}$ N{\smaller{V}}, C{\smaller{IV}}, He{\smaller{II}}) in their rest-UV spectra, or based on the presence of broad H$\alpha$ emission in their rest-frame optical spectra. Most of these also exhibit photometric signatures of AGN in the mid-IR from {\it Spitzer}/IRAC and MIPS observations (see Shapley et al. 2005).  As there are 46 galaxies with spectroscopic redshifts that place them within the protocluster (${\rm i.e.}$ with $z_{\rm spec}=$ 2.30 $\pm$ 0.04), this gives a protocluster BX/MD selected emission-line AGN fraction of $10.9_{-4.7}^{+7.4}$~per cent (1$\sigma$ errors calculated according to the prescription of Gehrels 1986).  To determine if this represents a significant enhancement over that of the field, we used the sample of $z_{\rm spec}>1.4$ emission-line AGN in GOODS-N presented by Reddy et al. (2006).  There are 188 BX/MD galaxies with $z_{\rm spec}>1.4$ in this field -- 4 of these are classed as AGN from the characteristic high-ionization emission lines found in their spectra.  This is a field fraction of $2.1_{-1.0}^{+1.7}$~per cent and so we find an enhancement of emission-line AGN in the protocluster by a factor $5.1_{-4.9}^{+7.5}$ (errors calculated following Barlow 2004).  Fisher's Exact (FE; Uitenbroek 1997) analysis was performed to determine the significance of this enhancement over the field - this technique indicates the result is significant at 98.4~per cent confidence.

\subsection{X-ray selected AGN}
To identify optical counterparts to the X-ray sources in HS 1700+64 we matched the X-ray source list to the $\cal R$--band catalogue of Steidel et al. (2004) using a likelihood ratio (LR) method (Sutherland \& Saunders 1992; L09).  To begin, any systematic offset between the X-ray and optical catalogues was accounted for by crudely matching the datasets using a 2~arcsec match radius.  We found small systematic offsets of $\delta$RA$=$ +0.03~arcsec and $\delta$Dec.$=$ +0.11~arcsec between the X-ray and optical catalogues; the X-ray source positions were subsequently shifted to match the optical images.

LRs were calculated for all candidate counterparts within a 3~arcsec radius of each X-ray source.  Optical counterparts with a threshold $LR>0.5$ were found for 86 X-ray sources (of the 134 with optical coverage) to $\cal R_{\rm AB}=$ 25.5, with an expected contamination rate of $\sim$8.5~per cent.  This sample of identifications was found to include 8 BX/MD selected galaxies, two of which have spectroscopic redshifts which place them in the protocluster (Table ~\ref{tab:properties}).

The X-ray sources were also identified with LAE/HAE counterparts using the same method.  There were 4 (2) X-ray sources identified with an LAE (HAE) counterpart having $LR>0.5$; the expected contamination rate of this sample is $\sim$3.5~per cent ($\sim$0.5~per cent) and so we expect all of these to be real identifications.  Two of the X-ray LAEs are spectroscopic protocluster members and are also emission-line AGN (BNB3 and BNB10).  One of the two X-ray detected HAEs (NB57) is a continuum selected object and is also one of the X-ray detected LAEs.  Consequently, there is a small overlap between the samples (see Table~\ref{tab:properties}).  One of the detected LAEs was found to be a QSO at low redshift and so we consider 4 X-ray emitting galaxies to be members of the protocluster.

The spectra for the X-ray sources were extracted using the {\footnotesize{ACIS EXTRACT}} IDL package (v.2008-03-04; Broos et al. 2002).  A description of the procedures used in {\footnotesize{ACIS EXTRACT}} can be found in Townsley et al. (2003) and Getman et al. (2005).  Source counts in each observation were extracted from the 95~per cent pixelated EEF PSF region (at 1.5 keV) and then summed to create the composite source spectrum.  In constructing the background spectrum, all point sources were first masked out using a circular region with a size 1.1x the 99~per cent PSF of the source (at 1.5 keV). A local region which enclosed a minimum of 150 background counts was then found for each observation -- these counts were extracted and combined by scaling the background counts in each observation by the ratio of the exposure in the source and background extraction regions.  These values were then summed over all observations to produce the composite background spectrum for each source.  The composite Area Response Files (ARFs) and Response Matrix Files (RMFs) were created by first using the standard {\footnotesize{CIAO}} tools {\footnotesize{MKARF}} and {\footnotesize{MKACISRMF}} to construct the ARFs and RMFs for each observation, respectively, then combining these using the {\footnotesize{FTOOLS ADDARF}} and {\footnotesize{ADDRMF}}, respectively, using the exposures to weight each observation.

The luminosities of all X-ray sources (except J170118.7$+$640837) were calculated using {\footnotesize{XSPEC}} (v12.3.0; Arnaud 1996) with the C-statistic (Cash 1979) assuming a power law with photon index $\Gamma=1.9$ (Nandra \& Pounds 1994), a Galactic line-of-sight H {\smaller{I}} column density of 2.28$\times10^{20}$~cm$^{-2}$ (Kalberla et al. 2005) and an intrinsic absorption component as a free parameter.  Due to the low number of counts (see Table~\ref{tab:properties}) the spectral fitting of J170118.7$+$640837 yielded only an upper limit on the 2-10 keV luminosity -- the full-band (0.5--10 keV) source photometry was consequently used to calculate this object's luminosity.  The 0.5--10 keV flux (calculated using Bayesian methodology, see L09 for details) was used together with an {\footnotesize{XSPEC}} model (with the above values for $\Gamma$ and the galactic absorption) to estimate the 2--10 keV X-ray luminosity, uncorrected for intrinsic absorption.  The uncertainties quoted in Table~\ref{tab:properties} are scaled from those of the 0.5--10 keV flux.

\subsubsection{BX/MD} 
Of the 8 X-ray sources identified with a BX/MD selected counterpart, we found that 5 were identified with a BX galaxy and 3 with an MD galaxy.  One of these X-ray detected MD galaxies is a spectroscopically confirmed protocluster member (MD157; $z=$ 2.293); this object has an unobscured 2--10~keV X-ray luminosity of $L_{\rm 2-10 keV}=$ 4.03$^{\rm +0.67}_{\rm -0.58}\times10^{44}$~erg~s$^{-1}$, and is also an emission-line AGN (see Shapley et al. 2005).  One of the 5 X-ray detected BX galaxies (BX116) is a bright (${\cal R}=21.65$), broad-line QSO (due to broad Ly$\alpha$, N{\smaller{V}}, Si{\smaller{IV}} and C{\smaller{IV}} emission lines in it's spectra) which also has a spectroscopic redshift that places it within the overdensity ($z=2.269$).  This object has an unobscured X-ray luminosity of $L_{\rm 2-10 keV}=$ 9.98$^{\rm +7.12}_{\rm -5.17}\times10^{43}$~erg~s$^{-1}$. The remaining 6 BX/MD X-ray detections have no spectroscopic information available.  Table~\ref{tab:properties} lists some basic information regarding the properties of the 4 X-ray detected galaxies that are thought to be associated with the $z=$ 2.3 protocluster (due to spectroscopic, Ly$\alpha$ or H$\alpha$ detections).

To assess whether any enhancement of X-ray selected AGN activity in the protocluster BX/MD sample is observed, a field sample of BX/MD galaxies was constructed from the X-ray detected BX/MD fraction in the Groth Westphal Strip (GWS; 200~ks) and the GOODS-N region of the {\it Chandra} Deep Field--North (CDF-N; 2Ms).  There are 1700 BX/MD galaxies in these fields, of which 9 are X-ray detected.  To ensure that we are taking into account the spatially varying X-ray detection limit characteristic of any {\it Chandra} observation, we constructed sensitivity maps for HS 1700+64 according to the methods described in Georgakakis et al. (2008b).  Due to the presence of the two low redshift X-ray clusters known to be in the HS 1700+64 field we excluded two elliptical regions around these objects from our analysis (see Fig.~\ref{fig:LAEdist}).  This was done as determining the sensitivity of our observations in the vicinity of these clusters becomes problematic due to the high local background.  The ellipses were centred on the clusters and total an area of $\sim$25.3 sq arcmin; the exclusion of this area results in 17 BX/MD galaxies being removed from the subsequent analysis.  We used the positions of the remaining 29 spectroscopically confirmed BX/MD protocluster galaxies in the full-band (0.5--10 keV) sensitivity, background and exposure maps together with an {\footnotesize{XSPEC}} model (consisting of a power law with slope $\Gamma=$ 1.4 and the above value for the Galactic absorption), to determine that the protocluster X-ray BX/MD AGN fraction is $6.9_{-4.4}^{+9.2}$~per cent for AGN with $L_{\rm 2-10 keV}\geq4.6\times10^{43}$~erg~s$^{-1}$ at $z=2.30$.  To calculate the corresponding AGN fraction we would expect to find in the protocluster based on the field sample, we summed up the numbers of field AGN that we would be sensitive to at each protocluster galaxy position and found, on average, we would expect to detect a BX/MD AGN fraction of $0.35_{-0.14}^{+0.21}$~per cent.  This means we would expect to detect $\sim$0.10 BX/MD selected AGN in the protocluster based on the field.  When this is compared to the 2 AGN that we actually observe, we find that this represents an enhancement of protocluster BX/MD X-ray AGN over the field by a factor of $19.9_{-4.4}^{+9.2}$ -- the FE method quantifies this result as significant at 99.3~per cent confidence.  We note that if some fraction of the 6 X-ray BX/MD galaxies without spectroscopic information are also located within the overdensity then this would imply a more significant result.

Regarding the overlap between the protocluster AGN detected via their X-ray emission and those classified via emission lines in their optical/near-IR spectra, we find that the two X-ray detected BX/MD protocluster galaxies (MD157 and BX116) are both also identified as AGN from their spectral emission lines (Table~\ref{tab:properties}).  The remaining three emission-line AGN are not found to be associated with X-ray sources using the likelihood ratio technique.

\subsubsection{Ly$\alpha$ and H$\alpha$} 
Using the deep Ly$\alpha$ (rest wavelength 1216$\rm \AA$) narrow-band imaging of the HS 1700+64 field, we find that of the 123 objects showing a Ly$\alpha$ excess, 4 are associated with X-ray sources.  Of these four, one is a QSO at $z=1.567$ and two are the BX/MD objects noted above (MD157 and BX116), which are both spectroscopically confirmed protocluster members.  The remaining LAE has no spectroscopic information available -- the unobscured X-ray luminosity for this object was calculated using the Ly$\alpha$ redshift.  As $\sim$88~per cent of the spectroscopic LAEs were found to lie at the protocluster redshift and both MD157 (=BNB10) and BX116 (=BNB3) are confirmed protocluster members, we assume that the third LAE (BNB110) is also a protocluster member.  The three X-ray emitting LAEs associated with the protocluster are shown in Fig.~\ref{fig:LAEdist}.

The field sample used to assess any enhancement of X-ray LAEs in the protocluster was that from the $z=$ 3.1 blank field survey of the Extended {\it Chandra} Deep Field-South (E-CDFS; Gronwall et al. 2007) -- this comparison sample was also utilised by Lehmer et al. (2009) to study the X-ray emitting LAEs in the $z=$ 3.09 protocluster in SSA 22.  Excluding the aforementioned regions around the foreground clusters from the analysis results in the removal of 17 LAEs from our sample, leaving 106 objects.  Of these, 3 have confirmed redshifts which lie outside the overdensity and are also removed from the protocluster LAE sample.  We then used the sensitivity information at the position of each LAE to calculate that the fraction of protocluster LAEs harbouring an X-ray AGN is $2.9_{-1.6}^{+2.9}$~per cent for AGN with $L_{\rm 2-10 keV}\geq4.6\times10^{43}$~erg~s$^{-1}$ at $z=2.30$.  To make fair comparisons to the higher redshift LAEs, the  narrow-band detection limit of the ECDF-S sample (${\rm NB}<25.4$) was applied to our $z=$ 2.3 sample, accounting for the redshift difference between the two. This results in 62 LAEs in which we could detect an AGN with the above X-ray luminosity, with 2 of these LAEs being detected in the X-ray data and which also satisfy the narrow-band detection limit.  There are 259 LAEs in the Gronwall et al. (2007) field sample of which 257 are covered by the {\it Chandra} data (Lehmer et al. 2005; Luo et al. 2008).  Two of the ECDF-S LAEs are detected -- one of these was a low redshift ($z=$ 1.6) AGN (see Lehmer et al. 2009) and was removed from the sample.  The other LAEs (including the X-ray source) are assumed to lie at $z=$ 3.1 due to previous spectroscopic observations (Gawiser et al. 2006; Gronwall et al. 2007).  Sensitivity maps for the ECDF-S were used as in $\S$3.2.1 to determine that we would be sensitive to AGN with $L_{\rm 2-10 keV}\geq4.6\times10^{43}$~erg~s$^{-1}$ (at $z=$ 3.1) in 212 of the 256 LAEs.  Comparing these protocluster (2/62) and field (1/212) X-ray LAE populations yields an enhancement of $6.8^{+4.3}_{-2.2}$.  When the FE analysis is applied this result is found to be significant at only 87~per cent confidence, which we do not consider to be indicative of a significantly enhanced population of X-ray protocluster LAEs over the field population.

We also note that of the 6 LABs in our Ly$\alpha$ narrow-band selected sample, none were found to be associated with X-ray sources.  Applying our LAB selection criteria ({\rm i.e.} $\rm EW_{obs}>90\rm \AA$, NB $<$ 25.5 and isophotal Ly$\alpha$ areas $>$50~arcsec$^2$) to the 29 LABs with {\it Chandra} coverage in the SSA 22 protocluster (Matsuda et al. 2004; Geach et al. 2009), results in a sample of 5 LABs (their LABs 1, 2, 3, 4 and 5).  Two of these are detected in the X-ray data (LAB2 and LAB3) giving an X-ray detected LAB fraction of 2/5.  Based on the X-ray fluxes we are sensitive to at the positions of our 6 LABs (using the methods outlined in the previous section) and the fluxes of LABs 2 and 3 from Geach et al. (2009), we would expect that (on average) we would detect 2 of the 6 LABs in our X-ray data.  The FE method quantifies the lack of a detection in HS 1700 as being consistent with the results from SSA 22.

Regarding the 62 H$\alpha$ narrow-band objects in this field, we find that two are detected in the X-ray data.  One of these is MD157 (Table~\ref{tab:properties}) whilst the other has an estimated X-ray luminosity of $L_{\rm 2-10 keV}=1.23^{+0.70}_{-0.55}\times 10^{43}$erg~s$^{-1}$, uncorrected for intrinsic absorption (see $\S 3.2$).  This object is also classified as a Distant Red Galaxy (DRG), characterised by having near-infrared colours that satisfy $J-K_{s}>$ 2.3 (Franx et al. 2003).  Due to the lack of an appropriate field comparison sample we are unable to quantify whether or not these detections represent an enhancement over the field population.

\begin{table*}
\begin{center}
\caption{Properties of Protocluster AGN. 
(1): Source R.A. (if no optical position then n.b. position is adopted); (2): Source Declination; (3): BX/MD continuum-selected ID (from Steidel et al. 2004); (4): X-ray ID (J2000); (5): Ly$\alpha$ ID; (6): H$\alpha$ ID; (7): Spectroscopic Classification (`QSO' denotes objects with emission features broader than 2000~km~s$^{-1}$); (8): Redshift; (9): Total X-ray counts (0.5--7 keV); (10): 2--10 keV band unobscured X-ray luminosity (uncertainties are 1$\sigma$ and scaled from those of the 2--10 keV flux)}
% {\smaller{
\begin{tabular}{cccccccccc}
\hline
R.A. & Dec. & BX/MD ID & X-ray ID & Ly$\alpha$ NB ID & H$\alpha$ NB ID & Spectroscopic & $z$ & Cts & $L_{\rm 2-10 keV}$ \\
(J2000) & (J2000) & & (CXO) &   &  & Class & & & ($\times10^{43}$~erg~s$^{-1}$) \\
(1) & (2) & (3) & (4) & (5) & (6) & (7) & (8) & (9) & (10) \\
\hline 
 255.136999 & 64.090158 & BX116 & J170032.9$+$640525 & BNB3 & - & QSO & 2.269 & 34 & $9.98_{-5.17} ^{+7.12}$ \\
 255.217612 & 64.258058 & $^{\it a}$MD157 & J170052.2$+$641529 & BNB10 & NB57  & AGN & 2.293 & 213 & $40.29_{-5.77} ^{+6.68}$ \\
 255.327842 & 64.143885 & - & J170118.7$+$640837 & - & NB6 & - & $^{\it c}$2.3 & 11 & $^{\it d}1.23^{+0.70}_{-0.55}$ \\
 255.481296 & 64.234314 & - & J170155.4$+$641402 & BNB110  & - & - & $^{\it c}$2.3 & 39 & $3.85_{-0.54} ^{+0.47}$ \\
 255.198546 & 64.162372 & $^{\it b}$MD69 & - & - & NB16 & AGN & 2.286 & - & - \\
 255.175234 & 64.189997 & $^{\it a}$MD94 & - & BNB165 & - & AGN & 2.333 & - & - \\
 255.227411 & 64.273479 & $^{\it a}$MD174 & - & - & - & AGN & 2.338 & - & - \\
\hline
\label{tab:properties}
\end{tabular}
%}}
\end{center}
\vspace{-0.5cm}
\begin{flushleft}
$^{\it a}$Further details on the AGN identification of these objects can be found in Shapley et al. (2005).\\
$^{\it b}$Hints of AGN activity are observed in this object's near-IR and optical spectra. He {\smaller{II}} and N {\smaller{II}} emission is observed, together with a ratio $\rm{[N{\smaller{II}}]/H\alpha} \sim 0.5$ and excess 8$\micron$ emission (see Shapley et al. 2005).\\
$^{\it c}$Assumed redshift for luminosity calculation.\\
$^{\it d}$The luminosity for this object was calculated from the source photometry, not from the source spectrum, and is uncorrected for intrinsic absorption. Uncertainties are scaled from those of the 0.5-10 keV flux.  See $\S 3.2$ for details.
\end{flushleft}
\end{table*}

\section{Summary and Discussion}
We have presented the results of spectroscopic, narrow-band and {\it Chandra} observations of a $z=$ 2.30 protocluster in the field of the QSO HS 1700+643.  We find evidence for an enhancement of emission-line BX/MD selected AGN in this protocluster, significant at $\sim$98.5~per cent confidence when compared to a similarly selected field sample from GOODS-North.  We also find evidence for an enhancement of X-ray detected BX/MD galaxies, significant at the $\sim$99~per cent confidence level, using a field sample constructed from the GWS and CDF-N.  Using a control sample of $z=3.1$ LAEs from the ECDF-S, we find that the X-ray emitting LAE fraction in the HS 1700 protocluster is not found to be significantly enhanced over that of the field (result significant at only 87~per cent confidence).

Considering the overlap between the emission-line and X-ray selected AGN in the BX/MD sample, we note that there are five emission-line AGN which are members of the protocluster, with two of these detected in the X-ray data.  This may seem disconcerting as X-rays are generally very penetrating and can overcome obscuring columns of $\lesssim$10$^{24}$~cm$^{-2}$ (although for column densities above this the source becomes Compton-Thick to X-rays) -- we may therefore expect a higher fraction of the emission-line AGN to be detected at X-ray wavelengths.  In the GOODS-N field however, Reddy et al. (2006, Table 5) found 4 BX/MD selected galaxies that were classified as AGN based on emission lines in their optical spectra.  Of these, two ({\rm i.e.} 50~per cent) had an X-ray counterpart within 1.5~arcsec in the 2Ms CDF-N data (Alexander et al. 2003), completely consistent with our fraction (2/5) within the errors.  This could indicate that the emission-line AGN are highly obscured or that the X-ray and emission-line AGN are two independent populations.  Recent studies of ten X-ray selected groups and six clusters at $z<0.06$ by Arnold et al. (2009) also find that the emission-line AGN population seems to be disconnected from the X-ray selected population - out of the 14 emission-line and six X-ray selected AGN in their study, only one was detected in both the X-ray data and via emission-line diagnostics.  Other results which seem to support this include X-ray observations of clusters finding a significantly larger number of AGN than identifications made via optical emission-line diagnostic methods (e.g. Martini et al. 2006), whilst optically-selected groups of galaxies are found to harbour a higher fraction of emission-line AGN compared to X-ray selected AGN ({\rm e.g.} Shen et al. 2007).  Arnold et al. (2009) suggest that the apparant disjointed nature of these two populations could reflect differing accretion modes for the emission-line (thin-disk accretion) and X-ray selected (radiatively-inefficient accretion) AGN; they claim that systems in various stages of virialization could have different dominant accretion modes and subsequently harbour a larger fraction of one type of AGN.  Within this framework, the non-virialized nature of the HS 1700 protocluster would result in a larger population of emission-line AGN being observed compared to that of the X-ray selected population -- the results from this study are consistent with this idea.

The results from HS 1700+64 can be compared to those of Lehmer et al. (2009), who examined the X-ray AGN content of a higher redshift protocluster at $z=$ 3.09 in the SSA 22 field.  They claim a suggested enhancement of X-ray LBGs with log $L_{\rm 8-32 keV}\gtrsim43.5$~erg~s$^{-1}$ is found in the protocluster, quoting a confidence level of $\sim$76~per cent.  They also claim a suggested excess of X-ray LAEs for protocluster AGN with log $L_{\rm 8-32 keV}\gtrsim 43.5$~erg~s$^{-1}$, quoting the result is significant at $\sim$78~per cent confidence.  When taken alone, these results do not represent convincing evidence for enhanced AGN fractions and it is only by combining the X-ray LBGs and LAEs that the authors inferred an excess AGN content at $\sim$95~per cent confidence, by a mean factor of $6.1^{+10.3}_{-3.6}$.  They attribute this to the presence of more actively growing and/or more massive SMBHs residing within the protocluster.  In the HS 1700+64 protocluster there is evidence for an enhancement of X-ray BX/MDs, by a factor of $19.9_{-4.4}^{+9.2}$, independently significant at $>$99~per cent confidence for AGN with $L_{\rm 2-10 keV}\geq4.6\times10^{43}$~erg~s$^{-1}$. Using the FE analysis, the factor by which the fraction of protocluster X-ray LAEs is greater than that of the field ($6.8^{+4.3}_{-2.2}$) is found only to be significant at 87~per cent confidence which, similar to the LAE result found for SSA 22, we do not consider to be indicative of evidence for an enhanced X-ray LAE population.  The simplest explanation for any observed excess of AGN activity in these protoclusters is that they contain more massive galaxies than the field.  Galaxies within the SSA 22 protocluster have inferred stellar masses a factor $\geq$ $1.2-1.8$ greater than field galaxies (Lehmer et al. 2009), and are $\sim$2 times as large for the HS 1700+64 protocluster (S05). Protocluster galaxies are also found to be significantly more evolved, on average, in terms of their stellar populations. At lower redshifts, X-ray emitting AGN are known to be hosted by massive, red galaxies ({\rm e.g.} Nandra et al. 2007; Bundy et al. 2008). Hence one explanation for the observed excess of AGN activity may simply be that protocluster galaxies have more massive and well developed bulges than field galaxies, and so we therefore observe a higher fraction of protocluster galaxies that exhibit AGN activity.  

With regard to the X-ray emitting LAE population in the HS 1700+64 protocluster, we found that when making a fair comparison to the $z=$ 3.1 ECDF-S field sample, the fraction of LAEs which harbour an AGN with $L_{\rm 2-10 keV}\geq4.6\times10^{43}$~erg~s$^{-1}$ in HS 1700 (2/62) is only significant at 87~per cent confidence when compared to that of the field (1/212).  We do not consider this result to represent evidence for an enhanced X-ray LAE fraction.  We do note however, that this field LAE sample is at a much greater redshift than the HS 1700+64 protocluster and so may not be representative of the LAE population at $z=$ 2.3.  Indeed, recent studies of LAEs as a function of redshift have revealed significant evolution in the intrinsic properties of these galaxies between $z\sim 3$ and $z=2.25$ (Nilsson et al. 2009).  Notably, the LAE population at $z=2.25$ in the Nilsson et al. (2009) study has a larger fraction of massive, old and/or dusty galaxies than at $z\sim3$, with the authors concluding that AGN are more numerous and also contribute more to the LAE population.  Recent results from Guaita et al. (2010), who studied a sample of 250 $z=2.1$ LAEs in the ECDF-S, also support a higher AGN fraction being present at these redshifts with the authors noting that the number density of X-ray detected AGN in their sample is consistent with that of Nilsson et al. (2009), and that their corresponding X-ray AGN fractions are also consistent when restricting their sample to the same Ly$\alpha$ luminosity limit.  We would therefore expect the $z\sim2$ X-ray LAE field fraction to be larger than at $z\sim 3$, and consequently would expect the fraction of X-ray LAEs in HS 1700+64 to be even less significant when compared to the field, meaning that our conclusion that the X-ray LAE fraction in HS 1700 is not significantly enhanced over the field would still hold.  Futhermore, we do not observe any X-ray emitting LABs in HS 1700, unlike the situation in the higher redshift protocluster in SSA 22 (Geach et al. 2009).  If AGN activity was indeed exciting the Ly$\alpha$ emitting population due to the dense environment at $z\sim 2$, then perhaps one would expect to detect some fraction of the LABs in the X-ray data.  We do note, however, that of the sample of 5 LABs from Geach et al. (2009) that fulfil our selection criteria, the two X-ray detected objects are at least factor of two brighter in Ly$\alpha$ luminosity than the LABs in our sample, and that the results are consistent within the errors.

To test the idea of mergers/galaxy interactions being the primary triggering mechanism behind the observed AGN activity, we used the {\it HST}/ACS I-band (F814W) imaging available in this field (Peter et al. 2007) to search for evidence of recent or ongoing merger activity.  It must be borne in mind, however, that at $z=2.3$ we are probing the rest-frame UV morphology of the galaxies with the F814W filter, not the optical morphology -- this means we are sensitive to the brightest star-forming regions of galaxies and not to the overall mass distribution.  Four AGN from Table~\ref{tab:properties} lie on the ACS footprint in the HS 1700+64 field -- MD157, MD174, MD69 and MD94.  Three of these (MD157, MD69 and MD94) exhibit disturbed morphologies with multiple nuclei and extended emission.  It is difficult to determine whether these characteristics are an indication of past merger events or galaxy interactions -- galaxies at $z>2$ are known to be highly irregular and not as easily classified into Hubble types as their local counterparts are.  As a result, even obviously disturbed morphologies cannot be directly linked to major mergers (see Law et al. 2007 and references therein).  Quantative morphological comparisons to the non-AGN population in these high-$z$ protoclusters are needed to determine if the distribution of AGN host galaxy morphologies is statistically different to that of the inactive protocluster galaxies and whether these differences, if any, are related to the incidence of mergers/galaxy interactions -- this work is beyond the scope of this paper.  We also note that Peter et al. (2007) did not find any evidence for the morphology of the general BX/MD galaxy population being a function of environment in this field -- this is unlike the situation at low redshift where morphology is a function of the galaxy number density (Dressler 1980), and where disturbed galaxies are usually indicative of a recent/ongoing merger event.  This also indicates that attributing specific processes ({\rm e.g.} galaxy interactions) to $z>2$ galaxies based on their morphology should only be done in an extremely cautious and quantative manner, not just by the (apparantly) disturbed nature of the galaxy's morphology.

Finally, in view of these small number statistics and those in the Lehmer et al. (2009) study, observations of other high-$z$ protoclusters are also required to constrain the nature of the AGN population in these structures and the implications for galaxy and cluster formation.

\section*{Acknowledgements}
We thank the anonymous referee for their comments which significantly improved this paper.  JADN and ESL acknowledge generous financial support from the STFC.  KN acknowledges the support of the Royal Society.  This work is based on observations taken with the Chandra X-Ray Observatory, which is operated for NASA by the Smithsonian Astrophysical Observatory.

\bibliographystyle{mn2e}
{}

\label{lastpage}

\end{document}